\documentclass[twocolumn]{aastex631}%
\usepackage{amsmath}

\begin{document}
\defcitealias{Liu2019ApJS..241...32L}{Paper\,I}
\defcitealias{Xiang2022A&A...662A..66X}{Xiang2022}
\title{The Value-added Catalog of OB Stars in LAMOST DR7}

\author[0000-0001-5314-2924]{Zhicun Liu}
\altaffiliation{Physics Postdoctoral Research Station at Hebei Normal University}
\affiliation{Department of Physics, Hebei Normal University, Shijiazhuang 050024, People's Republic of China}
\affiliation{Guo Shoujing Institute for Astronomy, Hebei Normal University, Shijiazhuang 050024, People's Republic of China}
\affiliation{Hebei Key Laboratory of Photophysics Research and Application, Shijiazhuang 050024, People's Republic of China}

\author[0000-0003-1359-9908]{Wenyuan Cui}
\affiliation{Department of Physics, Hebei Normal University, Shijiazhuang 050024, People's Republic of China}
\affiliation{Guo Shoujing Institute for Astronomy, Hebei Normal University, Shijiazhuang 050024, People's Republic of China}

\author{Jiajia Gu}
\affiliation{Department of Physics, Hebei Normal University, Shijiazhuang 050024, People's Republic of China}

\author[0000-0002-0349-7839]{Jianrong Shi}
\affiliation{CAS Key Laboratory of Optical Astronomy, National Astronomical Observatories, Chinese Academy of Sciences, Beijing 100101, People's Republic of China}

\author[0000-0003-1828-5318]{Guozhen Hu}
\altaffiliation{Physics Postdoctoral Research Station at Hebei Normal University}
\affiliation{Department of Physics, Hebei Normal University, Shijiazhuang 050024, People's Republic of China}
\affiliation{Guo Shoujing Institute for Astronomy, Hebei Normal University, Shijiazhuang 050024, People's Republic of China}
\affiliation{Hebei Key Laboratory of Photophysics Research and Application, Shijiazhuang 050024, People's Republic of China}

\author[0000-0003-2536-3142]{Xiao-Long Wang}
\altaffiliation{Physics Postdoctoral Research Station at Hebei Normal University}
\affiliation{Department of Physics, Hebei Normal University, Shijiazhuang 050024, People's Republic of China}
\affiliation{Guo Shoujing Institute for Astronomy, Hebei Normal University, Shijiazhuang 050024, People's Republic of China}
\affiliation{Hebei Key Laboratory of Photophysics Research and Application, Shijiazhuang 050024, People's Republic of China}

\author{Zhenyan Huo}
\affiliation{Department of Physics, Hebei Normal University, Shijiazhuang 050024, People's Republic of China}
\affiliation{Guo Shoujing Institute for Astronomy, Hebei Normal University, Shijiazhuang 050024, People's Republic of China}

\correspondingauthor{Wenyuan Cui}
\email{wenyuancui@126.com; cuiwenyuan@hebtu.edu.cn}






\begin{abstract}
In this work, we update the catalog of OB stars based on the Large Sky Area Multi-Object Fiber Spectroscopic Telescope (LAMOST) data release 7 and modified the OB stars selection criterion in spectral line indices’ space. The new catalog includes 37,778 spectra of 27,643 OB stars, of which 3827 OB stars are newly identified. The spectral subclasses of 27,643 OB stars are obtained using the automatic classification code MKCLASS. We find that the modified OB star selection criteria can better improve the completeness of late B-type stars by analyzing their spectral classification results given by MKCLASS. We also identify 3006 Be-type stars or candidates by examining the Balmer lines in their spectra and find that the frequency of our Be-type stars (10.9\%) is consistent with previous results. The spatial distribution of OB stars indicates that they are mainly located in the Galactic disk. This new catalog of OB stars will provide valuable data for studying the structure and evolution of the Milky Way.  
\end{abstract}




\section{Introduction} \label{sec:intro}

Massive stars have a significant impact on the chemical composition and evolution of host galaxies via their powerful winds, radiation fields, and supernova explosions \citep{Heger2003ApJ...591..288H,Evan2005A&A...437..467E,Evan2011A&A...530A.108E,Gray2009ssc..book.....G}. OB stars, which refer to any type of O and B stars in this work, are young stars with high effective temperatures and luminosities\citep[hereafter Xiang2022]{Morgan1973ARA&A..11...29M,Xiang2022A&A...662A..66X}. These characteristics make them suitable tracers for exploring the spiral arm structure, star-forming region, chemical evolution, and present-day cosmic abundances in the Milky Way \citep{Morgan1953ApJ...118..318M,Przybilla2012A&A...539A.143N,Xu2021A&A...645L...8X}.

In the last few decades, a series of large-scale survey projects have been implemented and gathered thousands of OB stars. For instance, to study the effects of mass loss, metallicity, rotation, and binary on the evolution of massive stars, the Very Large Telescope (VLT)-Fibre Large Array Multi-Element Spectrograph (FLAMES) survey has obtained the high-resolution spectra of over 700 massive OB stars\footnote{Here, massive OB stars refer to O and early B-type stars (formally B2 or earlier for dwarfs, B5 or earlier for giants, and all B subtypes for supergiants).}in the Galaxy and Magellanic Clouds\citep{Evan2005A&A...437..467E,Evan2006A&A...456..623E}, and the multi-epoch optical spectroscopy of over 800 massive stars in the 30 Doradus region of the Large Magellanic Cloud \citep{Evan2011A&A...530A.108E}. \citet{Maiz2011hsa6.conf..467M} presented spectra of over 1000 Galactic O-type stars with high signal-to-noise ratio and intermediate resolution (R\,$\sim$\,2500), based on the Galactic O-Star Spectroscopic Survey (GOSSS). Subsequently, \citet{Sota2011ApJS..193...24S,Sota2014ApJS..211...10S} and \citet{Maiz2016ApJS..224....4M} completed their spectral classification and improved the classification criteria for O-type stars. The IACOB spectroscopic database has compiled a homogeneous set of high-quality, high-resolution spectra of Galactic O- and early-B type stars using the high-resolution Fibre-fed Echelle Spectrograph\citep[FIES]{Simon2011BSRSL..80..514S,Simon2011sca..conf..255S,Simon2023A&A...674A.212D}.  \citet[hereafter Paper I] {Liu2019ApJS..241...32L} selected about 16,000 OB stars from the LAMOST DR5 low-resolution spectra using the line-indices method.

The large sample of OB stars with spectra data is important for the study of their stellar parameters and chemical composition. The projected rotational velocities ($v$\,sin\,$i$) of B-type stars in the cluster are systematically larger than those in the field by studying the distribution of $v$\,sin\,$i$ of a large sample of OB stars \citep{Dufton2006A&A...457..265D,Wolff2008AJ....136.1049W}.  \citet{Simon2014A&A...562A.135S} further confirmed that macroturbulent broadening is common in the domain of massive stars by studying the distribution of $v$\,sin\,$i$ of about 200 Galactic O- and early B-type stars. The study of the chemical composition of about 285 early B-type stars in the Galaxy and Magellanic Clouds indicates that nitrogen abundances can be used as a suitable indicator to investigate the rotational mixing of massive stars \citep{Hunter2009A&A...496..841H}. Recently, \citet{Simon2024A&A...687A.228D} has completed the large-scale quantitative spectroscopic analysis of 527 Galactic supergiants with spectral types from O9 to B5, and found that the relative numbers of stars and the numbers of rapid rotating stars decreased significantly when the stellar effective temperature was below 21000\,K.

Furthermore, the large sample of OB stars can also be used to study the structures of the Galactic disk and the distribution of Galactic spiral arms. \citet{Carraro2015AJ....149...12C} indicated that the flare of the Galactic thin disk can also be revealed using their OB star samples. Subsequently, \citet{Li2019ApJ...871..208L} used 250,000 OB stars from the \textit{Gaia} Data Release2 (\textit{Gaia} DR2) \citep{Gaia2016A&A...595A...2G} and the Two Micron
All-Sky Survey (2MASS) \citep{Skrutskie2006AJ....131.1163S} photometric catalogs to obtain the scale length of the Galactic disk, and found that the disk scale height increases with Galactic radius. Using the LAMOST DR5 13,534 OB stars with Galactocentric distance ranging from 8 to 14\,kpc, \citet{Yu2021ApJ...922...80Y} found that the flaring structure is symmetrical about the Galactic plane. \citet{Chen2019MNRAS.487.1400C} found that the Galactic spiral structure shows flocculent patterns, based on 14,880 O and early B-type stars and the kinematics parameters in the \textit{Gaia} DR2. \citet{Xu2021A&A...645L...8X} further investigated the local spiral structure using the \textit{Gaia} EDR3 9750 O-B2 stars with parallax accuracy better than 10\%, and their results revealed a clear sketch of nearby spiral arms in the Galaxy, especially in the third and fourth quadrants that absent the maser parallax data. 

\begin{table}
    \footnotesize
    \centering
    \caption{The definition of line indices}
    \begin{tabular}{lcc}
    \hline
    \hline
     Name&    Index Bandpass &Pseudocontinua \\
     &(\AA)&(\AA)\\
    \hline
    \ion{Ca}{2}~K &3927.700–3939.700 &3903.000$–$3923.000 4000.000$–$4020.000\\
    H$_\gamma$ & 4319.750–4363.500&4283.500–4319.750 4367.250–4419.750\\
    Fe4383&4370.375–4421.625&4360.375–4371.625 4444.125–4456.625\\
    Fe4531&4515.500–4560.500&4505.500–4515.500 4561.750–4580.500\\
    Fe4668&4635.250–4721.500&4612.750–4631.500 4744.000–4757.750\\
    Fe5015&4977.750–5054.000&4946.500–4977.750 5054.000–5065.250\\
    Fe5270&5245.650–5285.650&5233.150–5248.150 5285.650–5318.150\\
    Fe5335&5312.125–5352.125&5304.625–5315.875 5353.375–5363.375\\
    Fe5406&5387.500–5415.000&5376.250–5387.500 5415.000–5425.000\\
    Fe5709&5698.375–5722.125&5674.625–5698.375 5724.625–5738.375\\
    Fe5782&5778.375–5798.375&5767.125–5777.125 5799.625–5813.375\\
    \hline
    \end{tabular}
    \label{tab1:linedices}
\end{table}

More OB stars with spectra are crucial for studying massive star evolution, Galactic spiral arms, star-forming regions, etc. In this work, we aim to update the catalog of spectroscopic OB stars based on LAMOST DR7 data. The paper is organized as follows: In Section~\ref{DataMethod}, we describe the LAMSOT DR7 data and the method used to select OB stars. The results and discussion are shown in Section~\ref{RandD}. A summary is presented in Section~\ref{Summary}.

\section{Data and Method} \label{DataMethod}
\subsection{LAMOST Data}

LAMOST is a 4\,m, quasi-meridian, reflecting Schmidt telescope, which includes 4000 fibers in a field of view of 20 square degrees in the sky \citep{2012RAA....12.1197C,2012RAA....12..735D,2012RAA....12..723Z}. LAMOST DR7 has collected more than 10 million low-resolution spectra with a wavelength range from 3690 to 9100\,\AA, including more than 9 million stellar spectra.

\subsection{Line Indices}

\begin{figure}[htp!]
    \centering
    \includegraphics[width=9.0cm]{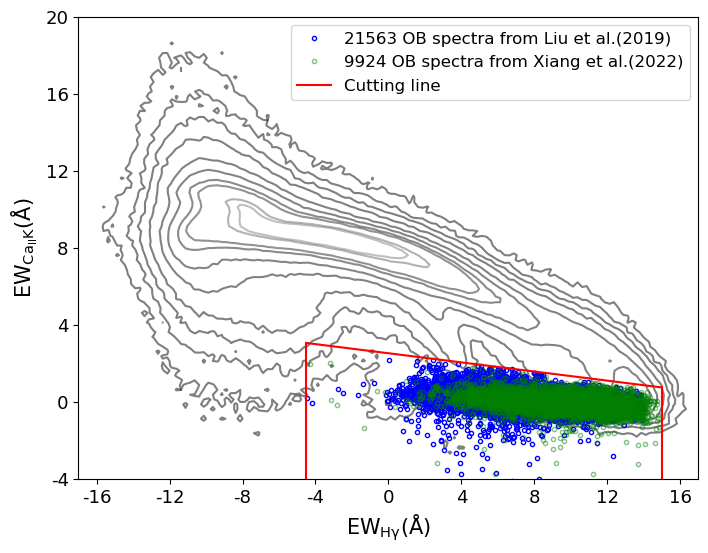}
    \caption{The distribution of LAMOST DR7 5,731,921 spectra with S$/$N$_g$\,$\ge$\,15 (gray contours), 21,529 OB spectra from \citetalias{Liu2019ApJS..241...32L} (blue dashed dots), and 9924 OB spectra selected from \citetalias{Xiang2022A&A...662A..66X} (green dashed dots) in the EW$_{\rm H\gamma}$ vs. EW$_{\rm CaIIK}$ plane. The red solid lines indicate the cuts used to remove contaminants.}
    \label{fig01:Figure01}
\end{figure}

The line indices in terms of equivalent width (EW) are deﬁned by the following equation \citep{Worthey1994ApJS...94..687W,Liu2015RAA....15.1137L}:

\begin{equation}\label{eq:ew}
\rm EW=\int(1-\frac{F_\lambda}{F_C})d\lambda,
\end{equation}
where F$_\lambda$ and F$_{\rm C}$ are the ﬂuxes of the spectral line and pseudo-continuum, respectively. F$_{\rm C}$ is estimated via linear interpolation of the ﬂuxes located in the “shoulder” region on either side of the line bandpass. The unit of the line index under this deﬁnition is in~\AA. 

As shown in \citetalias{Liu2019ApJS..241...32L}, the OB stars can be selected based on distributions in the spectral line indices’ space. Therefore, we calculate the EW of \ion{Ca}{2}\,K (3933\,\AA), H$_\gamma$ (4340\,\AA), and Fe line (EW$_{\rm Fe}$). Here, we denote EW$_{\rm Fe}$ as the mean value of Fe lines 4383, 4531, 4668, 5015, 5270, 5335, 5406, 5709, and 5782\,\AA. The deﬁnition of the line indices is listed in Table~\ref{tab1:linedices}.

\subsection{Method}\label{Method}

In \citetalias{Liu2019ApJS..241...32L}, we identified 22,901 OB spectra of 16,032 stars in LAMOST DR5 using spectral-line-indices method and final manual inspection, based on the LAMOST DR5 low-resolution spectra with signal-to-noise ratio larger than 15 at $g$\,band (S/N$_g$). However, the completeness of the OB spectra is only about 57$\pm$16 for the stars with spectral types later than B7. \citetalias{Xiang2022A&A...662A..66X} compiled 1,163,410 spectra of 844,790 hot-star candidates in LAMOST DR6 by combing the candidate OBA stars from \citet{2021A&A...650A.112Z}, OB stars from \citetalias{Liu2019ApJS..241...32L}, and OBA-type stars classified by LAMOST pipeline. They estimated the stellar parameters of about 330,000 OBA-type stars using the spectral fitting tool PAYNE.

\begin{figure}[htp!]
    \centering
    \includegraphics[width=9.0cm]{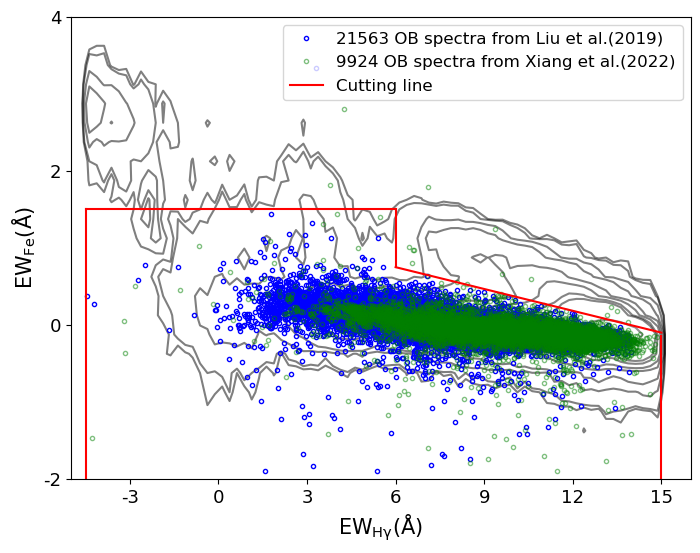}
    \caption{The distribution of LAMOST DR7 200,722 OB star candidates (gray contours), 21,529 OB spectra from \citetalias{Liu2019ApJS..241...32L} (blue dashed dots), and 9924 OB spectra selected from \citetalias{Xiang2022A&A...662A..66X} (green dashed dots) in the EW$_{\rm H\gamma}$ vs. EW$_{\rm Fe}$ plane. The red solid lines indicate the cuts used to remove contaminants.}
    \label{fig02:Figure02}
\end{figure}

In this work, our goal is to identify the “normal” OB stars in LAMOST DR7\footnote{Here, we deﬁne the “normal” OB stars that don't the some low-mass OB stars, such as hot subdwarfs, white dwarfs (WDs), blue horizontal-branch (BHB) stars, post-asymptotic giant branch (post-AGB) stars, etc.}. The 330,000 OBA-type stars of \citetalias{Xiang2022A&A...662A..66X} include OB stars, A-type stars, hot subdwarfs, white dwarfs, and so on. In order to better mark the distribution of known OB stars of \citetalias{Xiang2022A&A...662A..66X} in the spectral index space, we use the following steps to select OB stars from the 330,000 known OBA-type stars  

\begin{itemize}
\item[(i)] Removing the 22,901 OB spectra in \citetalias{Liu2019ApJS..241...32L}.
\item[(ii)] To facilitate manual inspection, we select the stars with spectral (S/N)$_g$\,$\geq$\,15.
\item[(iii)]Taking account of the uncertainties of the effective temperature and the effective-temperature range of OB stars, we select the stars with $T_{\rm eff}$ $\geq$ 9000\,K. The stellar $T_{\rm eff}$ is from \citetalias{Xiang2022A&A...662A..66X}.
\end{itemize}

After following the above steps, we identify 96,223 OB star spectra. We inspect the above spectra and identify 10,123 “normal” OB star spectra corresponding to 8085 OB stars. We also inspect 22901 OB spectra from \citetalias{Liu2019ApJS..241...32L} and obtain 21,669 “normal” OB star spectra corresponding to 16,138 OB stars. In addition, some stars have been observed multiple times and contribute several spectra. Considering the difference in EW given by different S/N spectra of the same star, we don't remove multiple spectra of the same stars for both the 10,123 and 21,669 OB spectra samples. 

In Figure~\ref{fig01:Figure01}, we map the 9924 of 10,123 OB spectra in \citetalias{Xiang2022A&A...662A..66X}, the 21,563 of 21,669 OB spectra in \citetalias{Liu2019ApJS..241...32L}\footnote{The 199 of 10,123 OB spectra don't have the \ion{Ca}{2}\,K, H$_\gamma$ or Fe measurements due to the bad fluxes in the relevant wavelength regions, and the 106 of 21,669 OB spectra don't have the \ion{Ca}{2}\,K, H$_\gamma$ or Fe measurements due to the bad fluxes in the relevant wavelength regions.}, and the 5,731,921 spectra with S$/$N$_g$\,$\ge$\,15 from LAMOST DR7 in the EW$_{\rm H\gamma}$ vs. EW$_{\rm CaK}$ plane, and find that our previous method of selecting OB stars in spectral line indices’ space underestimates the equivalent width of H$_{\rm \gamma}$. Therefore, we adopt the following empirical criteria to identify more OB stars in LAMOST DR7.  

\begin{equation}\label{equation2}
\begin{aligned}
    \rm EW_{\rm CaK}<2.5-EW_{\rm H\gamma}/8 \\
    \rm -4.5\,\leq\,EW_{\rm H\gamma}\,\leq\,15. \,
\end{aligned}
\end{equation} 

The 200,722 OB star candidates have been selected using the criteria described in the empirical equation~\ref{equation2}. As described in \citetalias{Liu2019ApJS..241...32L}, the distribution of OB stars in the Fe versus H$_\gamma$ plane can be utilized to remove most of the late-type stars. In Figure~\ref{fig02:Figure02}, we present the distribution of the 9924 OB star spectra from \citetalias{Xiang2022A&A...662A..66X}, 21,563 OB star spectra from \citetalias{Liu2019ApJS..241...32L}, and 200,722 spectra of OB star candidates in the EW$_{\rm H\gamma}$ vs. EW$_{\rm Fe}$ plane, and it is seen that most of the OB stars in \citetalias{Xiang2022A&A...662A..66X} are located in the region with larger EW$_{\rm H\gamma}$. We adopt the following empirical criteria in the EW$_{\rm H\gamma}$ vs. EW$_{\rm Fe}$ plane, so that most of the OB star candidates are included.

\begin{equation}\label{equation3}
\begin{aligned}
    \rm EW_{\rm Fe}<1.5\,and\,EW_{\rm H\gamma}<6 \\
    \rm EW_{\rm H\gamma}>6 \, and\, EW_{\rm Fe}<-0.095*EW_{\rm H\gamma}+1.316 \,
\end{aligned}
\end{equation} 


After applying the criteria~\ref{equation3}, the number of OB star spectra is reduced to 162,888. Additionally, there are 2496 spectra without \ion{Ca}{2}\,K, H$_\gamma$ or Fe measurements because these spectra have bad fluxes in the relevant wavelength regions. We inspect the spectra of 165,384 (16,288$+$2496) OB candidates and identify 37,719 OB spectra. 

\begin{figure}
    \centering
    \includegraphics[width=0.55\textwidth]{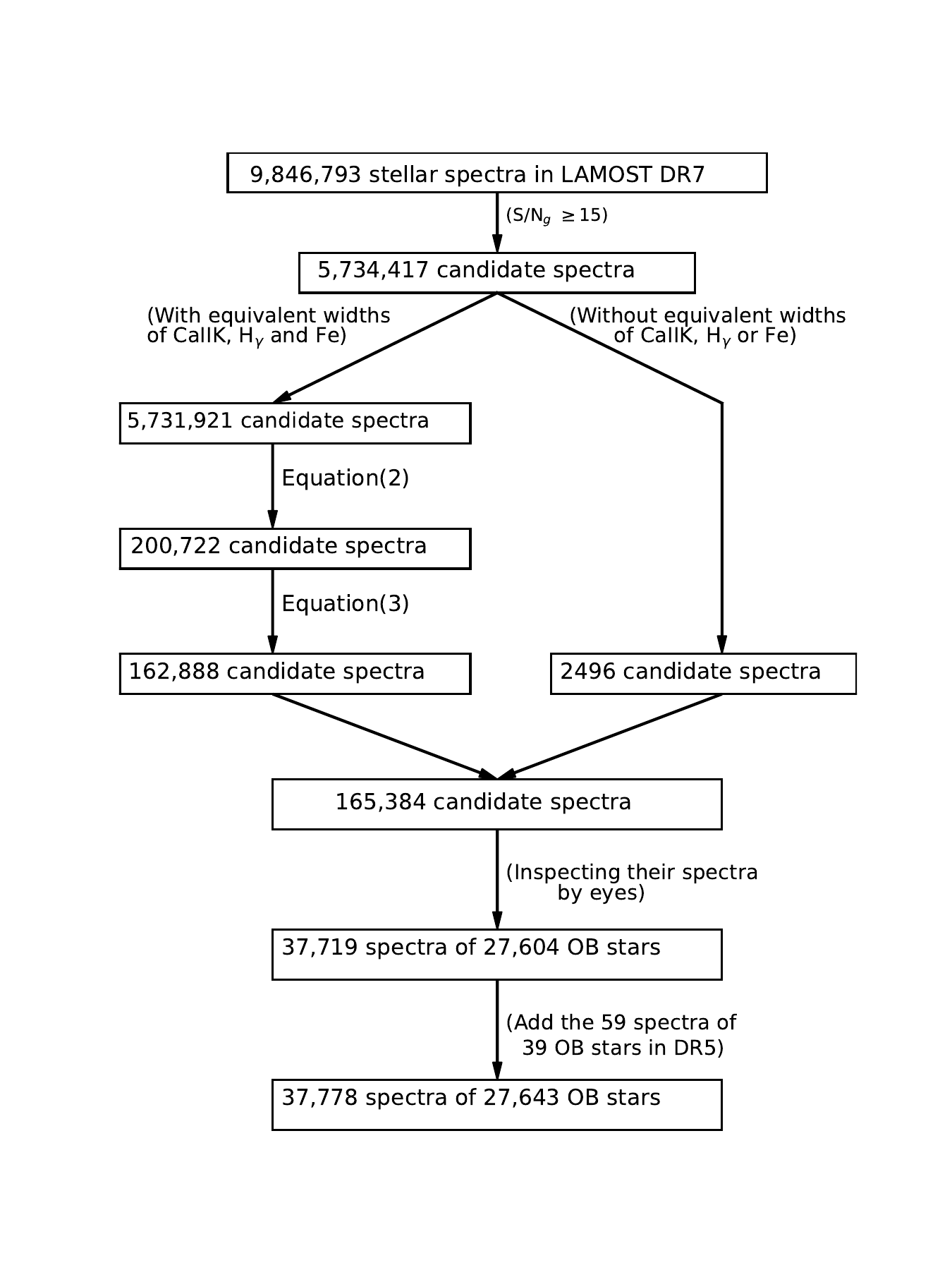}
    \caption{The steps of OB stars identification.}
    \label{figliucheng}
\end{figure}

In addition, we also find that 59 OB spectra don't be included in our OB stars samples by cross-matching 37,719 OB spectra in LAMOST DR7 and 21,669 OB spectra from \citetalias{Liu2019ApJS..241...32L}. This is because the LAMOST DR7 survey data does not fully include the DR5 survey data. Finally, the final sample of OB stars contains 37,778 (37,719 in DR7 +59 in DR5) spectra of 27,643 OB stars. The steps of identifying OB stars are summarized in Figure~\ref{figliucheng}. 

By cross-matching the above 27,643 OB stars with 16,138  “normal” OB stars from \citetalias{Liu2019ApJS..241...32L} and 8085  “normal” OB stars selected from \citetalias{Xiang2022A&A...662A..66X}\footnote{Here, we don't remove the common stars between 16,138 OB stars from \citetalias{Liu2019ApJS..241...32L} and 8085 OB stars selected from \citetalias{Xiang2022A&A...662A..66X}.}, Our 27,643 OB stars are described below: 

(i) There are 16,138 common OB stars with \citetalias{Liu2019ApJS..241...32L}.

(ii) There are 7678 common OB stars selected from \citetalias{Xiang2022A&A...662A..66X}.

(iii) 3827 OB stars are newly identified.

\begin{figure*}
    \centering
    \includegraphics[width=18.0cm,height=7.0cm]{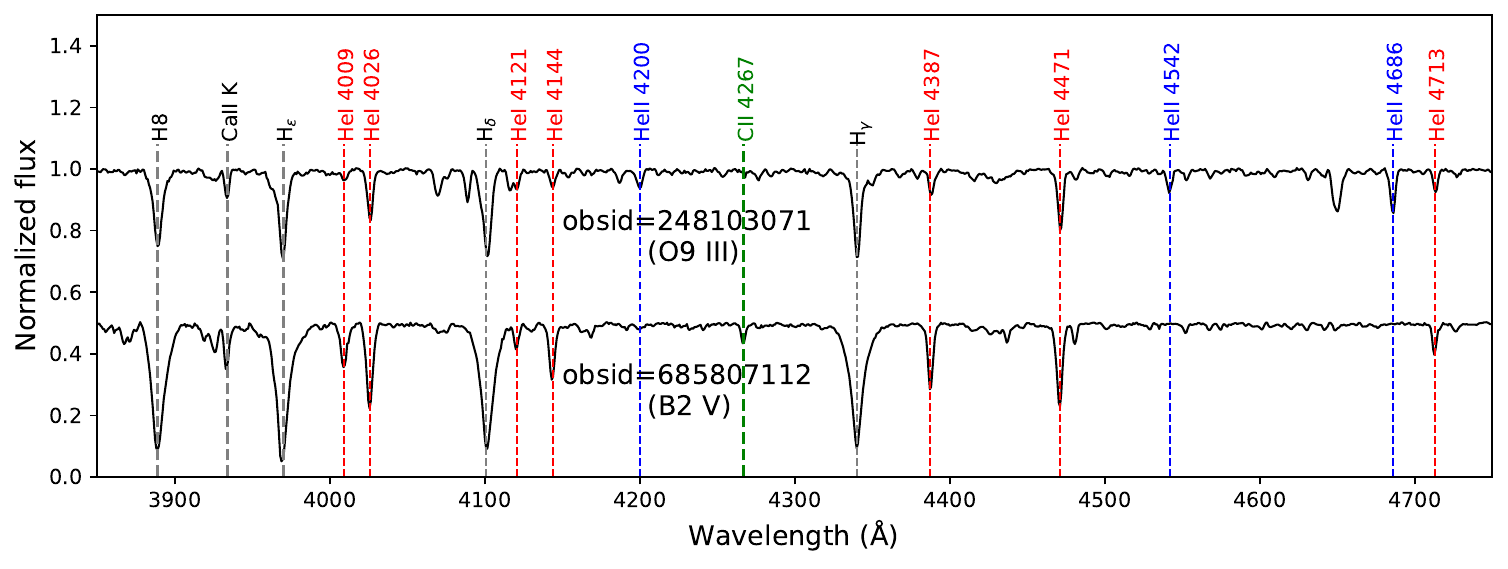}
    \caption{The example spectra of O (top) and B-type stars (bottom) in our LAMOST catalog, and some important line features, such as \ion{He}{1}\,4009, 4026, 4387, 4471, \ion{He}{2}\,4200, 4541, 4686\,\AA, \ion{C}{2}\,4267\,\AA, are also marked in the plots. Their spectral types obtained by manual classification are also shown in the picture}
    \label{fig03:Figure03}
\end{figure*}

\section{Results and Discussion} 
\label{RandD}
\subsection{spectral classification}

\begin{figure}
    \centering
    \includegraphics[width=0.45\textwidth]{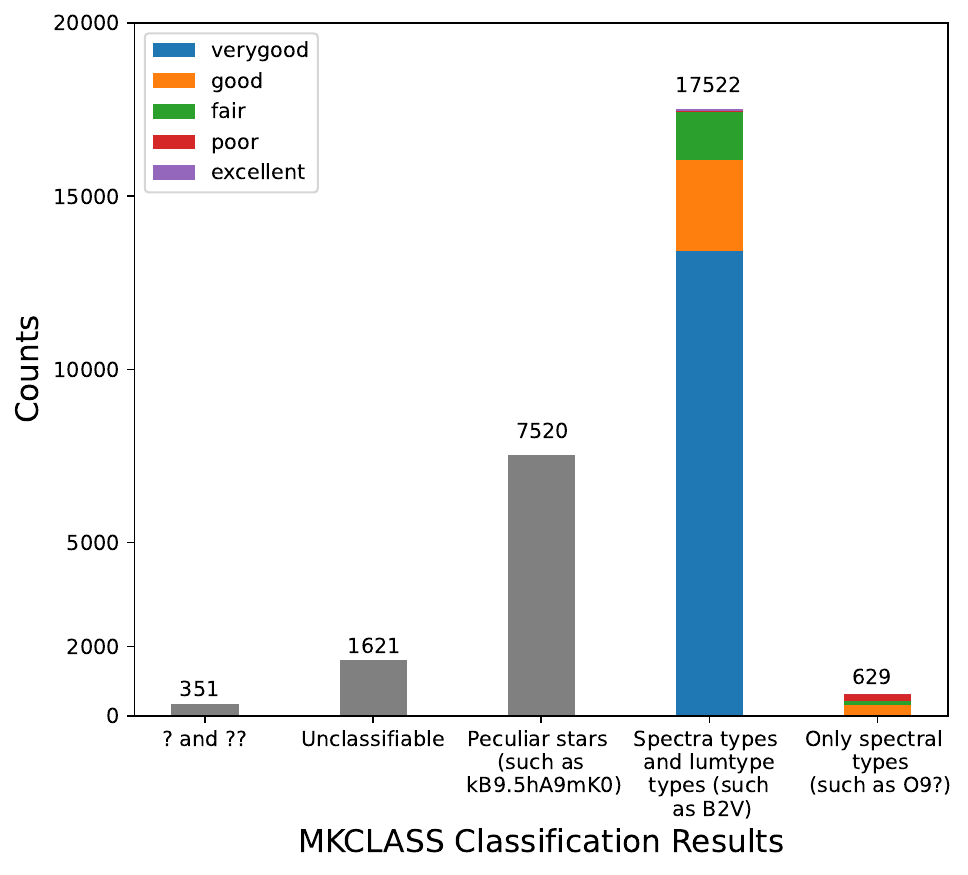}
    \caption{Histogram of the MKCLASS classification results of 27,643 OB stars. The performance of classification ('excellent', 'verygood', 'good', 'fair', and 'poor') given by MKCLASS are marked using different color bars for these stars with spectral types or luminosity types. }
    \label{fig04:Figure04}
\end{figure}

Stellar classification plays an important role in studying the stellar parameters \citep{1990PASP..102..379W,Gray2009ssc..book.....G,Sota2011ApJS..193...24S}. For example, the effective temperature and surface gravity of a star can be initially estimated from its spectral type and luminosity type, respectively. The main character of OB stars is the presence of absorption lines of \ion{He}{2} (O-type stars) and \ion{He}{1} (B-type stars) at their blue$–$violet spectra.

In Figure \ref{fig03:Figure03}, we show the blue spectra of one O9\,III O-type and one B2\,V B-type star in LAMOST DR7. Based on the spectral characters of OB stars and LAMOST low-resolution spectra, we check the 37,778 spectra of 27,643 OB stars and classify them as 279 spectra of 228 O-type stars and  37,499 spectra of 27,415 B-type stars. In addition, the 3827 newly identified OB stars include 27 O-type stars and 3800 B-type stars. The results are listed in Table \ref{Table2}.  

\subsubsection{Sub-classification of OB stars with MKCLASS}

The MKCLASS code, developed by \citet{2014AJ....147...80G}, is designed to classify blue$–$violet spectra in the MK spectral classification system by comparing the spectra of the program stars and MK standard stars \citep{1943assw.book.....M}. MKCLASS can provide spectral types for stars and assess its performance of classification using 'excellent', 'verygood', 'good', 'fair', and 'poor', based on the $\chi^2$ between the program one and the best-matched one of the standard star. There are two standard libraries in MKCLASS: (i) $libnor36$: the library includes MK standard star spectra with rectified blue$–$violet (3800$-$5600\,\AA) spectra of a resolving power R$\sim$ 1100, obtained with the Dark Sky Observatory. (ii) $libr18$: the library is constructed based on flux-calibrated spectra covering 3800$–$4600\,\AA \, with a resolving power R$\sim$2200. 

\begin{figure*}[htp!]
    \centering
    \includegraphics[width=16.5cm,height=7.5cm]{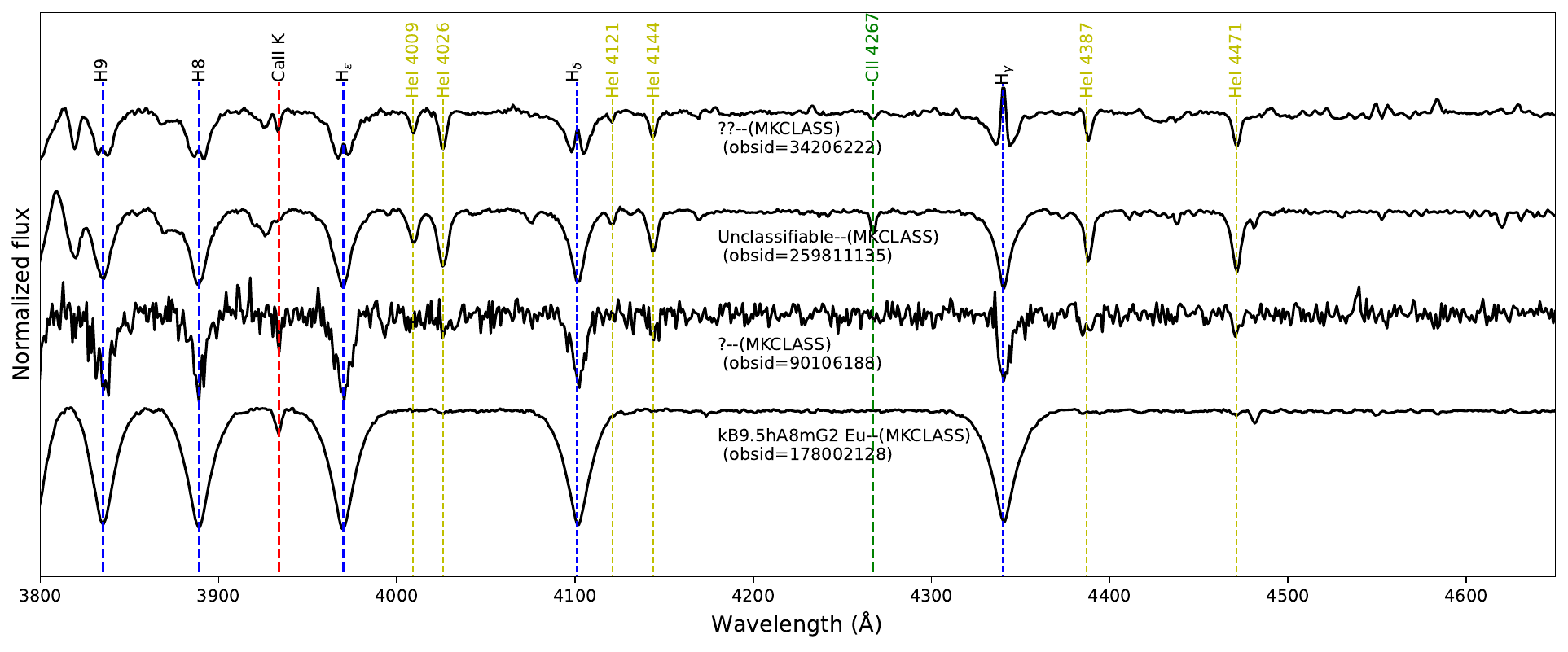}
    \caption{Sample of B-type spectra that the MKCLASS do not give a determined spectral type, such as Be-type star (first spectrum), He-Strong B-type star (second spectrum), B-type star with low S/N (third spectrum), and peculiar stars (fourth spectrum). Some important line features of B-type star spectra are also marked in the picture.}
    \label{fig06:Figure06}
\end{figure*}

\begin{figure*}[htp!]
    \centering
    \includegraphics[width=18.0cm,height=8.5cm]{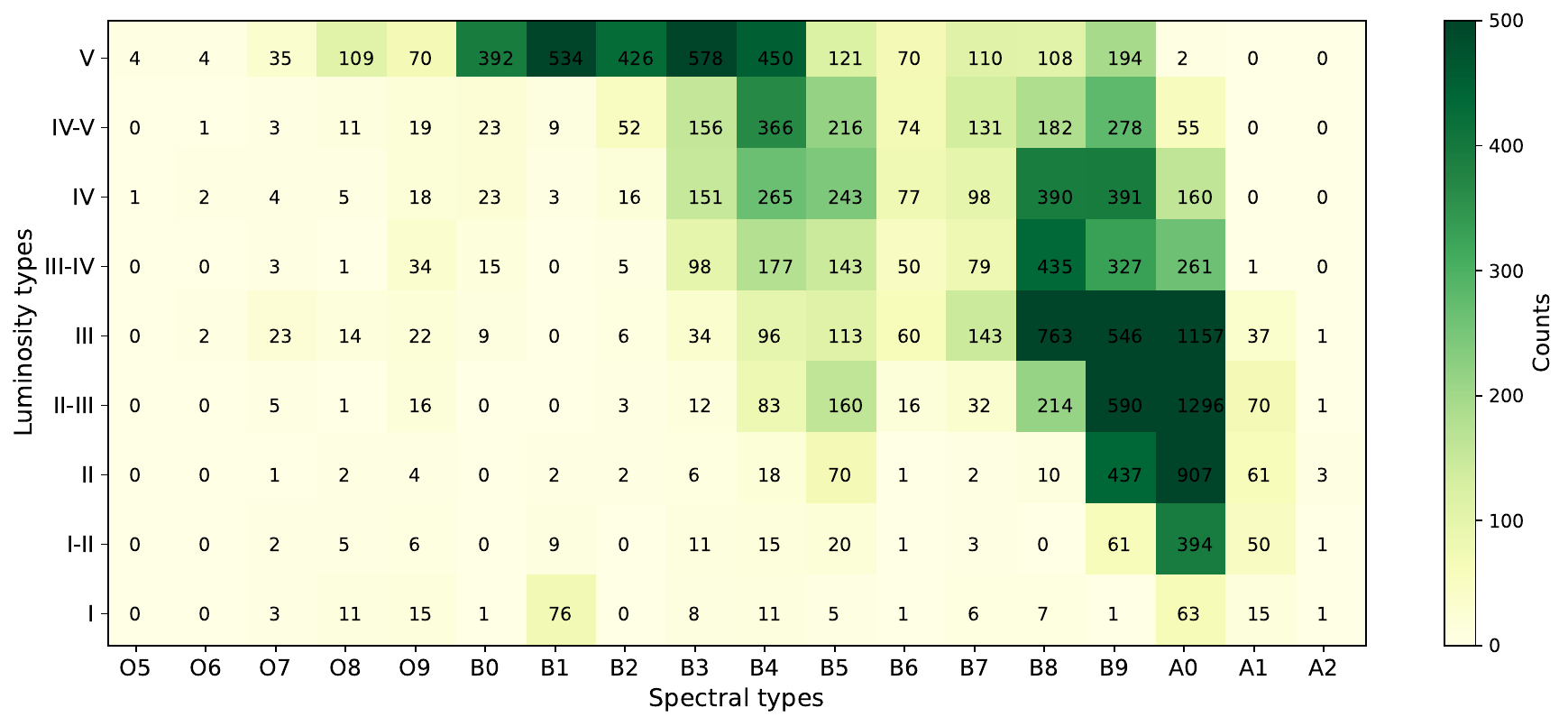}
    \caption{Distribution of 16,110 OB stars in the spectral subtype vs. luminosity class plane. The spectral types of these stars are obtained using the MKCLASS with quality evaluation better than “good” and including “good.”. The numbers in the different color boxes represent the number of stars in this spectral type range.}
    \label{fig05:Figure05}
\end{figure*}

Because the resolution of the LAMOST low-resolution spectra is about 1800 corresponding to a full wave at half maximum (FWHM) of 2.8\,\AA, we reduce the resolution of the library $libr18$ from 1.8 to 2.8\,\AA\ and denote the new library as $libr$18$\_28$. The research results in \citetalias{Liu2019ApJS..241...32L} indicate that MKCLASS can give a reliable classification for normal stars, based on the $libr$18$\_28$ library and LAMOST low-resolution spectra. Consequently, we submitted the 37,778 spectra of 27,643 OB stars to MKCLASS and obtained their spectral types. To facilitate the analysis of the MKCLASS results of 27,643 OB stars, the spectral types are obtained by using the highest signal-to-noise ratio spectra for the stars with multiple visits.

\subsubsection{MKCLASS Classification Results}\label{section312}

Figure~\ref{fig04:Figure04} shows the distribution of MKCLASS classification results for 27,643 OB stars. It is seen that only 1972 stars (7.1\%) cannot be classified by MKCLASS because their spectra have low S/N or emission lines after checking those spectra, and we also show sample spectra in Figure~\ref{fig06:Figure06}. For 7520 stars (27.2\%), MKCLASS cannot give a unique type based on their different characteristic lines, but discrepant classiﬁcations instead, such as: "kB9.5hA8mG2 Eu" which means that the K-line determined spectral type is B9.5, the hydrogen-line determined type is A8, the metallic-line classiﬁed type is G2, and this star may show strong Eu lines (see the fourth spectrum in Figure \ref{fig06:Figure06}). It also implies that LAMOST survey data contains many peculiar early-type stars \citepalias{Xiang2022A&A...662A..66X}. There are 18,151 stars (65.7\%) given spectral types by MKCLASS with different performance of classification.

In Figure~\ref{fig05:Figure05}, we show the distribution of 16,110 OB stars with the MKCLASS quality evaluation better than "good" and including "good" in spectral types versus the luminosity class plane. It is seen that the stars with a spectral type earlier than B6 are mostly main-sequence stars (luminosity classes IV-V and V), and most of the stars with a spectral type between B8 and A1 are giants (II-III). The distributions of OB stars in the spectral types versus luminosity class plane also suggest that the OB stars in the LAMOST survey are mainly late-type B dwarfs, which may be due to the observational selection effect. In addition, the distribution of supergiant stars (I-II and I) of different spectral types shows two peaks between spectral types B1 and B9-A0. The former peak may be due to the "blue loops" in massive stars \citep{Fitzpatrick1990ApJ...363..119F,Schaller1992A&AS...96..269S,Georgy2012A&A...537A.146E,Wagle2019ApJ...886...27W}, while the later peak is probably due to sample selection effects.

\subsection{Comparison with other works}

\subsubsection{Cross-match with Li (2021)}

\citet{Li2021ApJS..253...54L} have presented the spectral classification of 209 O-type, based on the LAMOST low-resolution spectral data between 2011 and 2018 using the spectral classification criteria for O-type stars summarized from \citet{Sota2011ApJS..193...24S,Sota2014ApJS..211...10S,Maiz2016ApJS..224....4M}. We cross-match the O-type star catalog of \citet{Li2021ApJS..253...54L} with the LAMOST DR7 catalog, and get 102 common O-type stars. The reason that the remaining 107 O-type stars of \citet{Li2021ApJS..253...54L} are not included in the LAMOST DR7 catalog as follows: most of the O-type stars in \citet{Li2021ApJS..253...54L} are collected from the LAMOST test sky and not released (\citep[private communication]{Li2021ApJS..253...54L} ). We cross-match the 102 O-type stars with our OB star catalog, and find they are included in our OB star samples.

\begin{figure}
    \centering
    \includegraphics[width=0.52\textwidth]{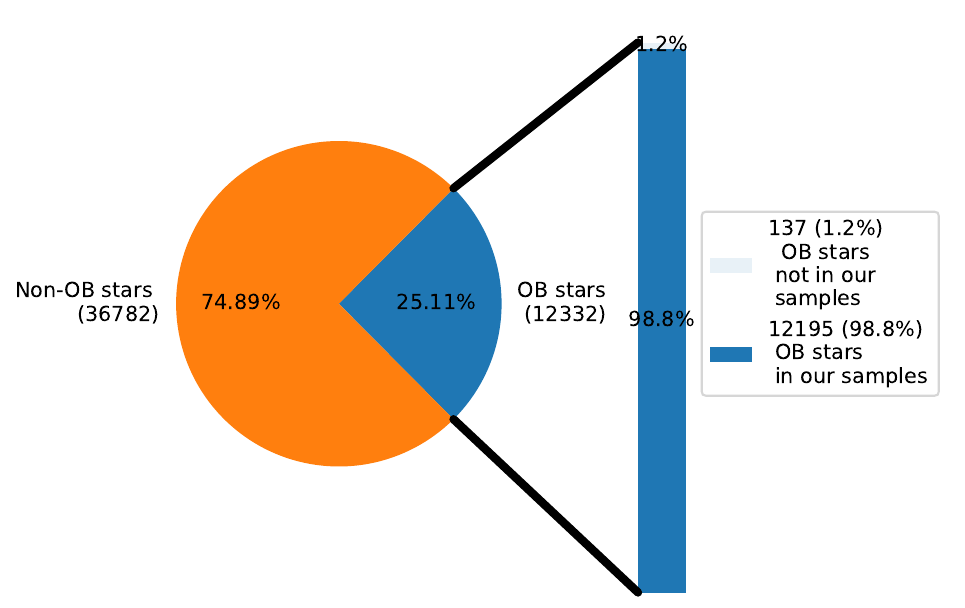}
    \caption{Left pie graph: the proportions are shown for 36,782 non-OB stars (74.89\%) and 12,332 OB stars (25.11\%) among 49,114 common stars. Right histogram: the ratio of 137 OB stars (1.2\%) not in our samples and 12,195 OB stars (98.8\%) in our samples.} \label{fig07:Figure07}
\end{figure}

\subsubsection{Cross-match with Zari et al.(2021)}

\begin{figure}[htp!]
    \centering
    \includegraphics[width=0.45\textwidth]{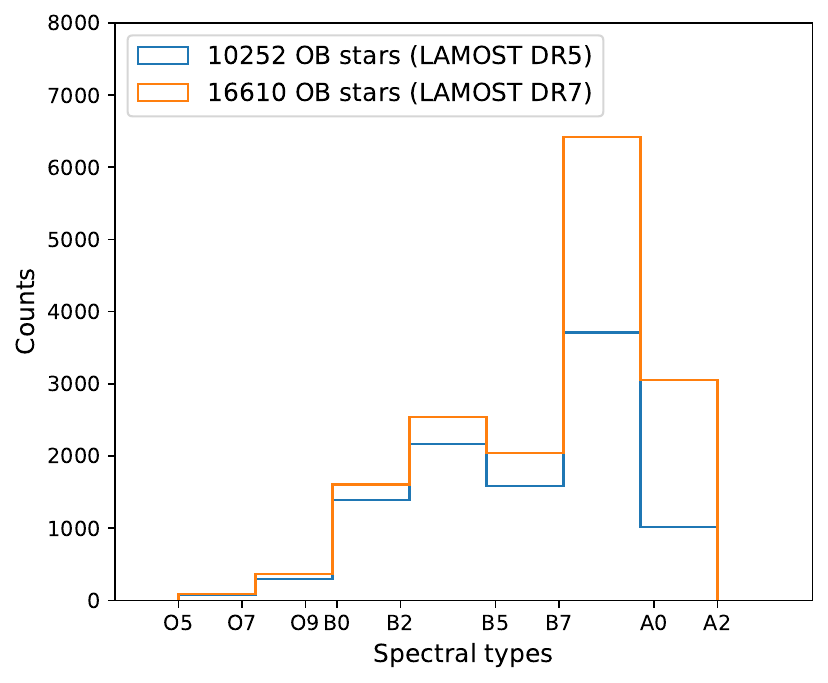}
    \caption{The distribution of spectral types of 10,252 and 16,610 OB stars in LAMOST DR5 and DR7, respectively.} \label{fig08:Figure08}
\end{figure}

\citet{2021A&A...650A.112Z} constructed an all-sky sample of 988,202 OBA-type stars, based on \textit{Gaia} EDR3 astrometry and 2MASS photometry. We cross-matched the above catalog with the LAMOST DR7 catalog and obtained 49,114 common stars. As shown in Figure~\ref{fig07:Figure07}, 12,332 of the them common stars are OB stars, and 36,782 are not OB stars. 74.89\% of the common stars are not OB stars because the catalog of \citet{2021A&A...650A.112Z} mainly contains A-type stars. It is also seen that 137 of the 12,332 OB stars are not included in our samples by cross-matching with our 27,643 OB stars. The reasons why we lost these 137 OB stars are as follows: (i)among them, 124 OB stars with spectral S/N$_{\rm g}$ $<$ 15 are removed due to our identification criterion, as the spectra of OB stars with S/N$_{\rm g}$ $<$ 15 are difficult to identify by the human inspect. (ii) 13 OB stars are missed due to the mistake \ion{Ca}{2}\,K, H$_\gamma$ or Fe measurements obtained by their bad spectra. 

\begin{figure*}[htp!]
    \centering
    \includegraphics[width=0.7\textwidth]{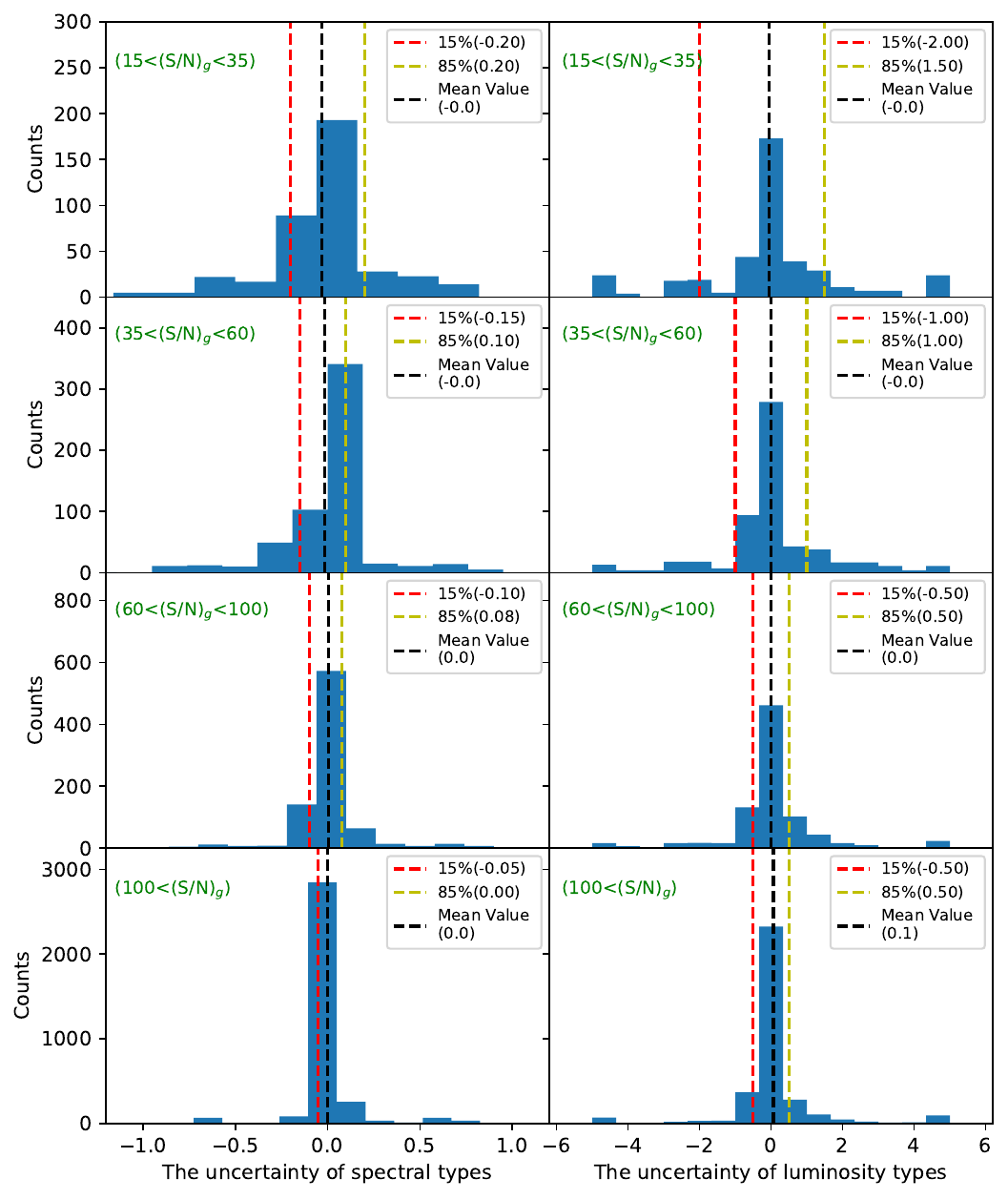}
    \caption{From top to bottom show the uncertainty distribution of the multiply observed OB stars with different spectral (S/N)$_g$ ranges in the spectral (left panel) and luminosity class (right panel). The mean value, 15\%, and 85\% percentile of spectral types uncertainty are marked using black, red, and red dashed lines in the left panel, respectively. The values of 15\% and 85\% percentile of spectral types uncertainty are also marked in parentheses (0.2 corresponding to 2 spectral subtypes). Similar to the left panel, the right panel shows the distribution of luminosity types uncertainty (2.0 corresponding to 2 luminosity subtypes).} 
    \label{fig000:Figure000}
\end{figure*}

\subsubsection{Comparison with \citetalias{Liu2019ApJS..241...32L} }

\begin{figure*}[htp!]
    \centering
    \includegraphics[width=20.5cm,height=7.8cm]{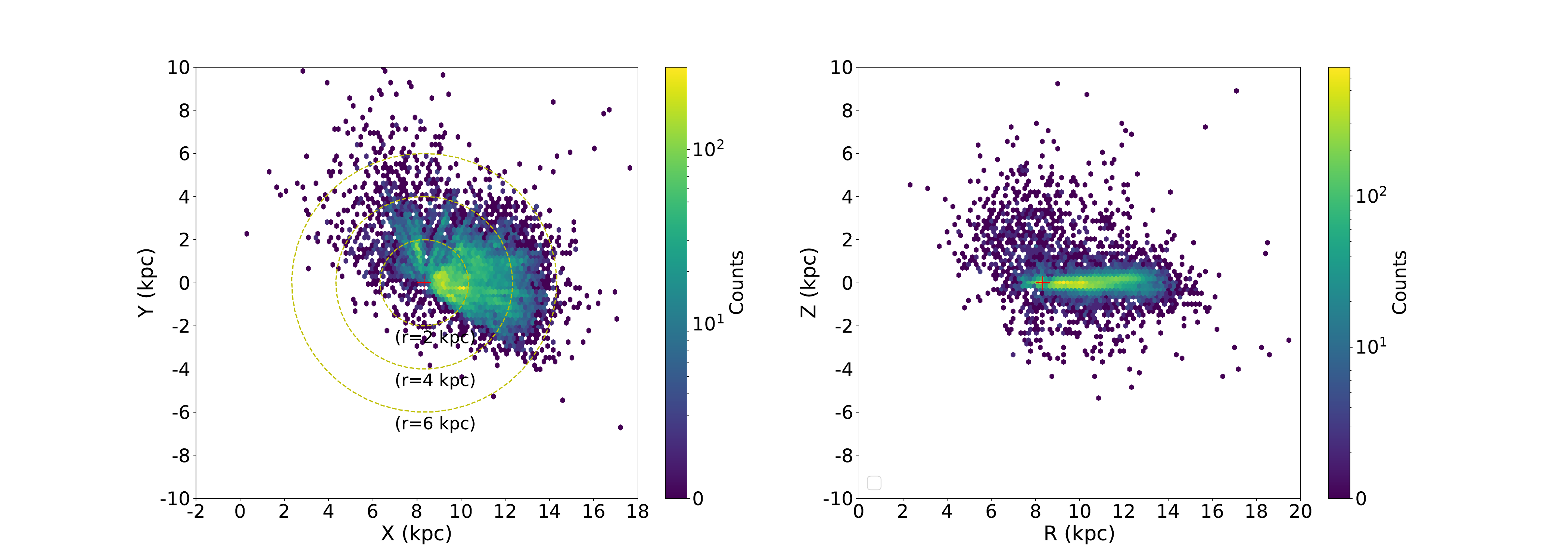}
    \caption{The distribution of 20,581 OB stars in the Galactic X-Y plane (left panel) and R-Z plane (right panel). The red cross represents the position of the Sun (X=8.33\,kpc, Y=0\,kpc, Z=0\,kpc). The yellow dashed rings delineate constant distances from the Sun in steps of 2\,kpc. The color bar corresponds to the number of stars.} 
    \label{fig09:Figure09}
\end{figure*}

As described in Section~\ref{section312}, the spectral types of 16,610 of 27,643 OB stars are obtained using the MKCLASS with a quality evaluation better than “good” and including “good.” After cross-matching 16,610 OB stars with the OB stars of \citetalias{Liu2019ApJS..241...32L}, we get 10,252 common OB stars. Figure~\ref{fig08:Figure08} shows the distribution of 16,610 OB stars in LAMOST DR7 and 10,252 OB stars in LAMOST DR5 in the spectral types plane. It is seen that OB stars identified in LAMOST DR7 include more late B-type stars and early A-type stars compared to those in LAMOST DR5. Meanwhile, \citetalias{Liu2019ApJS..241...32L} suggested that there is 1 subtype difference between the manual and MKCLASS classification results. Therefore, the stars with spectral types of A1 or A0 given by MKCLASS are actually B-type stars. Compared with the OB star identification method in \citetalias{Liu2019ApJS..241...32L}, we set higher equivalent widths for  H$_\gamma$ and Fe in the modified method, which will improves the completeness of late B-type stars. 

Given the (S/N)$_g$ and low-resolution of the LAMOST spectra, we estimate the uncertainty in the provided spectral and luminosity class for the multiply observed OB stars and show in Figure \ref{fig000:Figure000}. For convenience, we use the numbers to represent the spectral subtype and luminosity subtype, 0.1-0.9 corresponding to O1-O9, 1.0-1.9 corresponding to B0-B1, and 1-5 corresponding to I-V, such as spectral type=1.95 and luminosity=5 for the B9.5V. The spectral type uncertainty of the multiply observed OB stars is obtained using the spectral type obtained by their highest (S/N)$_g$ spectra as a criterion. For example, a star with three different (S/N)$_g$ spectra ((S/N)$_g$=20, (S/N)$_g$=50, and (S/N)$_g$=130), for the stars with spectral (S/N)$_g$=20 and (S/N)$_g$=50, the uncertainties are 0.1 and 0.0 for spectral types and 1 and 0 for luminosity types. As shown in Figure \ref{fig000:Figure000}, the uncertainties of spectral types are about 2, 1.5, and 1 subtype for the stars with spectral (S/N)$_g$$<$35, 35$<$(S/N)$_g$$<$60, and 60$<$(S/N)$_g$, respectively. while the uncertainties of luminosity types are about 2, 1, and 0.5 subtypes for the stars with spectral (S/N)$_g$$<$35, 35$<$(S/N)$_g$$<$60, and 60$<$(S/N)$_g$, respectively.

\subsection{Be-type stars}

Be-type stars, which are fast-rotating B-type main-sequence stars, display one or more Balmer emission lines in their spectra because they present an equatorial, circumstellar, accretion disk \citep{1987pbes.coll....3C,2003PASP..115.1153P,2013A&ARv..21...69R}.  The frequency of Be-type stars will increase with decreased metallicity. Variety studies indicate that approximately 10$–$20 percent of the B-type stars in our Galaxy are Be-type stars (11 percent in \citet{Jaschek1983A&A...117..357J}; 17 percent in \citet{1997A&A...318..443Z} ). There is a higher percentage of Be-type stars in the Large Magellanic Cloud (17-22\%) and the Small Magellanic Cloud (26-40\%)\citep{Martayan2006A&A...445..931M,Schootemeijer2022A&A...667A.100S,Martayan2007A&A...472..577M}.
We inspect the 37,499 spectra of 27,415 B-type stars and identify 3932 spectra of 3006 Be-type stars or candidates. The frequency of our Be-type stars is about 10.96$\%$. The reasons for the lower frequency of our Be-type stars are as follows: (i) most of our Be-type stars are field stars, and clusters have a higher ratio of Be-type stars than that of the field stars \citep{Keller2001AJ....122..248K,Yu2018AJ....155...91Y}. (ii) there is a lower fraction of Be-type stars in late-type B stars than in early-type B stars \citep{1997A&A...318..443Z,Porter2003PASP..115.1153P}. As shown in Figure~\ref{fig05:Figure05}, our B-type sample candidates are mainly late-type B stars. The large number of late B-type stars in the current sample makes the percentage of our Be-type star lower. 

\subsection{Spatial distribution of OB stars}

With \textit{Gaia} EDR3 data \citep{Gaia2023A&A...674A...1G}, we select 20,581 OB stars satisfying RUWE$\leq$1.4 and $\sigma_\omega$/$\omega\leq$30\%, and calculate their Galactocentric coordinates (X,Y,Z) by using the photogeometric distances from \citet{Bailer2021AJ....161..147B} and the solar position from \citet{Reid2014ApJ...783..130R} (R$_0$=8.34\,kpc). 

Figure~\ref{fig09:Figure09} shows the Galactocentric spatial distribution for our 20,581 OB stars. From the left panel of Figure~\ref{fig09:Figure09}, it is seen that the OB stars are mainly located in the 6\,kpc range of the Sun on the Galactic anti-center. From the right panel of Figure~\ref{fig09:Figure09}, it is seen that OB stars are mainly located in the Galactic disk. In addition, there are a small number of OB stars located in the Galactic plane with larger vertical distances (Z). These OB stars with larger Z, which may include some runaway OB stars, are very valuable for us to investigate the origins of runaway OB stars \citep{Hoggerwerf2001A&A...365...49H,Martin2004AJ....128.2474M,Martin2006AJ....131.3047M,Irrgang2022A&A...663A..39B,Liu2023MNRAS.519..995L}.

\section{Summary} \label{Summary}

In this work, we employ the modified spectral-line-indices method to update the catalog of OB stars based on the LAMOST DR7 data. The main results are summarized below:

(1) Based on the OB stars selected from \citetalias{Xiang2022A&A...662A..66X}, we modify the OB selection criteria in spectral line indices’ space. Using the revised selection criteria, we identify 37,778 spectra of 27,643 OB stars from the LAMOST DR7 low-resolution spectra. 3827 of them are newly identified.  

(2) We apply MKCLASS to classify the 27,643 OB stars, and find they are mainly B-type main-sequences with spectral types later than B2. It is found that the improved method can better identify OB stars in LAMOST survey data, especially for the late-B-type stars.

(3) There are 3932 spectra of 3006 Be-type stars or candidates identified by examining the Balmer emission lines in LAMOST spectra of B-type star sample, and the frequency (10.9\%) of our Be-type stars is consistent with previous results.

(4) The spatial distribution of OB stars suggests that most of them are located in the Galactic disk, which will provide important data support for investigating the Galactic structure in future work. In addition, there are some OB stars with large vertical distances in our samples, and the study of their origin provides a possibility to explore the formation mechanism of runaway OB stars \citep{Martin2004AJ....128.2474M,McEvoy2017ApJ...842...32M}.

\setcounter{secnumdepth}{-1} 
\section{Acknowledgments}.
We thank the anonymous referee for the helpful suggestions to help improve this manuscript. This study is supported by the National Natural Science Foundation of China under grants No. 12173013, 12090044, 12003045, and 12403034; the National Key Basic R$\&$D Program of China No. 2019YFA0405500; the project of Hebei provincial department of science and technology under the grant No. 226Z7604G, the Hebei NSFC (Nos. A2021205006), the project of Hebei Natural Science Foundation (No. A2024205031), and Science Foundation of Hebei Normal University(No. L2024B54, L2024B55, and L2024B56). The Guoshoujing Telescope (the Large Sky Area Multi-Object Fiber Spectroscopic Telescope LAMOST) is a National Major Scientific Project built by the Chinese Academy of Sciences. LAMOST is operated and managed by the National Astronomical Observatories, Chinese Academy of Sciences. This work has made use of data from the European Space Agency (ESA)  mission \textit{Gaia} (\url{https://www.cosmos.esa.int/gaia}), processed by the \textit{Gaia} Data Processing and Analysis Consortium (DPAC, \url{https://www.cosmos.esa.int/web/gaia/dpac/consortium}). Funding for the DPAC has been provided by national institutions, in particular the institutions participating in the Gaia Multilateral Agreement. This research has made use of the SIMBAD database, operated at CDS, Strasbourg, France. 

\begin{table*}[h]
\scriptsize
\caption{The basic information for 37,778 spectra of 27,643 OB stars identiﬁed in this work} \label{Table2}
\centering
\begingroup
\renewcommand{\arraystretch}{1.25}
\begin{tabular}{ccccccccc}
\hline
\hline
Obsid&Designation& \multicolumn{1}{c}{S/N$_{\rm g}^{a}$}&\multicolumn{1}{c}{MKCLASS}&\multicolumn{1}{c}{SIMBAD}& \multicolumn{1}{c}{Manual$^b$}&\multicolumn{1}{c}{Source ID}&\multicolumn{1}{c}{Parallax}&\multicolumn{1}{c}{Distances$^c$}\\
\cline{1-3} 
\cline{7-8}
\multicolumn{3}{c}{(LAMOST)}&\multicolumn{2}{c}{}&\multicolumn{1}{c}{}&\multicolumn{2}{c}{\textit{Gaia} EDR3}&\multicolumn{1}{c}{(pc)}\\
\hline
   100001177 & J045557.58+211459.1 & 164.3 & kA0hF1mG2 &  & B           & 3412127903494703616 & 0.8706$\pm$0.0297   & 1090.62$^{+34.12}_{-35.31}$\\
  100002184 & J044623.33+221203.5 & 198.01 & A0 II    & OB & B          & 3413030778636383232 & 0.4715 $\pm$ 0.0301 & 1857.13$^{+ 104.05 }_{-102.76}$\\
  100005136 & J045336.02+221209.6 & 70.75  & B9.5 IVn &            & B  & 3412423981361441152 &                     & 1592.16 $^{+ 33.22 }_{-39.15}$\\
  100005210 & J044946.51+215704.8 & 52.75  & A7 mA0 V metal-weak &  & B & 3412777268193287552 &                     & 2560.87$^{+ 275.46 }_{-197.61}$\\
  100007030 & J045604.80+213831.3 & 197.16 & A0 II &  & B               & 3412331622383534720 & 0.5586$\pm$0.0255   & 1632.9 $^{+ 69.11  }_{-65.25}$\\
  100007094 & J050147.68+214621.4 & 258.28 & kB9hA5mG8  Eu & A5 & B     & 3412196455470112384 & 1.1113$\pm$0.0204   & 876.25 $^{+14.19   }_{-15.15}$\\
  100008008 & J045630.77+223527.7 & 342.53 & A3 mA0 V Lam Boo & A0 & B  & 3412540259011937408 & 1.3717$\pm$0.0195   & 711.09 $^{+ 9.57   }_{-8.85}$\\
  100008050 & J045716.85+230328.4 & 77.47  & A0 II-III &  & B           & 3412592760691790592 &                     & 2579.57 $^{+ 115.15}_{-88.3}$\\
  100008248 & J045605.71+224956.5 & 208.05 & A1 IV & A2 & B             & 3412564482627163776 & 0.991$\pm$ 0.0389   & 957.69 $^{+ 43.25  }_{-30.63}$\\
  100013039 & J050338.55+234349.5 & 223.35 & A6 mB7 V Lam Boo &  & B    & 3418682062245934336 &                     & 687.01 $^{+ 20.85  }_{-18.86}$\\
  100013047 & J050415.97+234142.2 & 175.9 & A3 mA0 V Lam Boo & A & B    & 3418670822315071744 & 1.3519 $\pm$ 0.0184 & 716.65 $^{+ 9.92   }_{-8.28}$\\
  100013076 & J050210.96+233018.1 & 335.43 & B6 V & A0 & B              & 3418580632297530624 & 1.4397 $\pm$ 0.0187 & 679.11 $^{+8.84    }_{-7.55}$\\
  100013198 & J050338.41+240655.7 & 279.33 & A5 mB0 V Lam Boo & B9 & B  & 3418764869215028992 & 1.3879 $\pm$ 0.0184 & 704.52 $^{+9.81 }_{-11.16}$\\
  100014014 & J044413.65+241445.4 & 136.45 & B9 II-III & B9/A0: & B     & 147219079337102976 & 0.6444 $\pm$ 0.03    & 1464.54$^{+67.58 }_{- 62.63}$\\
  100014051 & J044655.69+233304.9 & 175.02 & Unclassifiable & F5 & B    & 146937363842973952 & 1.2331 $\pm$ 0.0227  & 784.07 $^{+11.56 }_{- 12.8}$\\
  100101061 & J045116.61+213003.7 & 375.32 & B9 V & A0 & B              & 3411998229842267264 & 1.6296 $\pm$ 0.0206 & 601.61 $^{+8.6 }_{- 7.36}$\\
  100105073 & J045106.60+223320.6 & 48.19 & B9.5 IV-V &  & B            & 3413242950020426240 & 0.5508 $\pm$0.0174  & 1690.26 $^{+36.5 }_{- 42.28}$\\
  100106018 & J050048.28+222400.3 & 42.73 & B8 V &  & B                 & 3412313484738734208 &                     & 1764.18 $^{+81.19  }_{- 60.53}$\\
  100106169 & J050321.11+224232.3 & 106.12 & B9 II-III &  & B           & 3418315443837401856 & 0.2978 $\pm$0.0202  & 2687.67 $^{+106.3  }_{- 114.26}$\\
  100106190 & J050103.07+223638.7 & 152.1 & B9 IVn & A0 & B             & 3412319978729250560 & 1.4726 $\pm$0.0269  & 656.07  $^{+10.99  }_{- 11.3}$\\
  ... & ... & ...& ... & ... & ... & ... & ...  & ...\\
  \hline
   \multicolumn{9}{l}{(a)The signal to noise at the g band.}\\
   \multicolumn{9}{l}{(b)The spectral type (O or B) identiﬁed by eye in this work.}\\
   \multicolumn{9}{l}{(c)The distances of OB stars are from \citet{Bailer2021AJ....161..147B}}\\
   \multicolumn{9}{l}{(This table is available in its entirety in machine-readable form.)}\\
 \end{tabular}
\endgroup
\end{table*}
%



\software{topcat \citep{Taylor2005ASPC..347...29T}, 
          laspec \citep{Zhang2020ApJS..246....9Z,Zhang2021ApJS..256...14Z}
          }








\bibliography{sample631}{}
\bibliographystyle{aasjournal}



\end{document}